\documentclass[aps,prl,reprint,superscriptaddress]{revtex4-1}
\usepackage{graphicx}
\usepackage{siunitx}
\usepackage{xcolor}
\usepackage{amsmath}
\usepackage{hyperref}
\hypersetup{
	colorlinks=true,
	linkcolor=blue,
	urlcolor=blue,
	citecolor=blue,
	pdfpagemode=FullScreen,
}
\usepackage[capitalise]{cleveref}

\usepackage{enumitem}
\usepackage{float}
\makeatletter
\def\maketitle{
	\@author@finish
	\title@column\titleblock@produce
	\suppressfloats[t]}
\makeatother

\begin{document}
	\title{Anomalous Hall Effect in Thin Bismuth}
	\author{Oulin Yu}
	\affiliation{Department of Physics, McGill University, Montréal, Québec, H3A 2A7, Canada}
	\author{Sujatha Vijayakrishnan}
	\affiliation{Department of Physics, McGill University, Montréal, Québec, H3A 2A7, Canada}
	\author{R. Allgayer}
	\affiliation{Department of Mining and Materials Engineering, McGill University, Montréal, Québec, H3A 2A7, Canada}
	\author{T. Szkopek}
	\affiliation{Department of Electrical and Computer Engineering, McGill University, Montréal, Québec, H3A 2A7, Canada}
	\author{G. Gervais}
	\affiliation{Department of Physics, McGill University, Montréal, Québec, H3A 2A7, Canada}
	\begin{abstract}
		Bismuth, the heaviest of all  group V elements with strong spin-orbit coupling, is famously known to exhibit many interesting transport properties, and effects such as Shubnikov-de Haas and de Haas-van Alphen were first revealed in its bulk form. However, the transport properties have not yet been fully explored experimentally in thin bismuth nor in its 2D limit. In this work, bismuth flakes with average thicknesses ranging from 29 to 69 nm were mechanically exfoliated by a micro-trench technique and were used to fabricate four-point devices. Due to mixing of  components, Onsager's relations were used to extract the  longitudinal ($R_{xx}$) and Hall ($R_{xy}$) resistances where the latter shows a Hall anomaly that is consistent with the Anomalous Hall Effect (AHE). Our work strongly suggests that that there could be a hidden mechanism for time-reversal symmetry breaking in pure bismuth thin films.
	\end{abstract}
	\date{\today}
	\pagenumbering{arabic}
	\maketitle
	
	\setlength{\parskip}{1em}
	
	\begin{figure}[tpb]
		\begingroup
		\centering
		\includegraphics[]{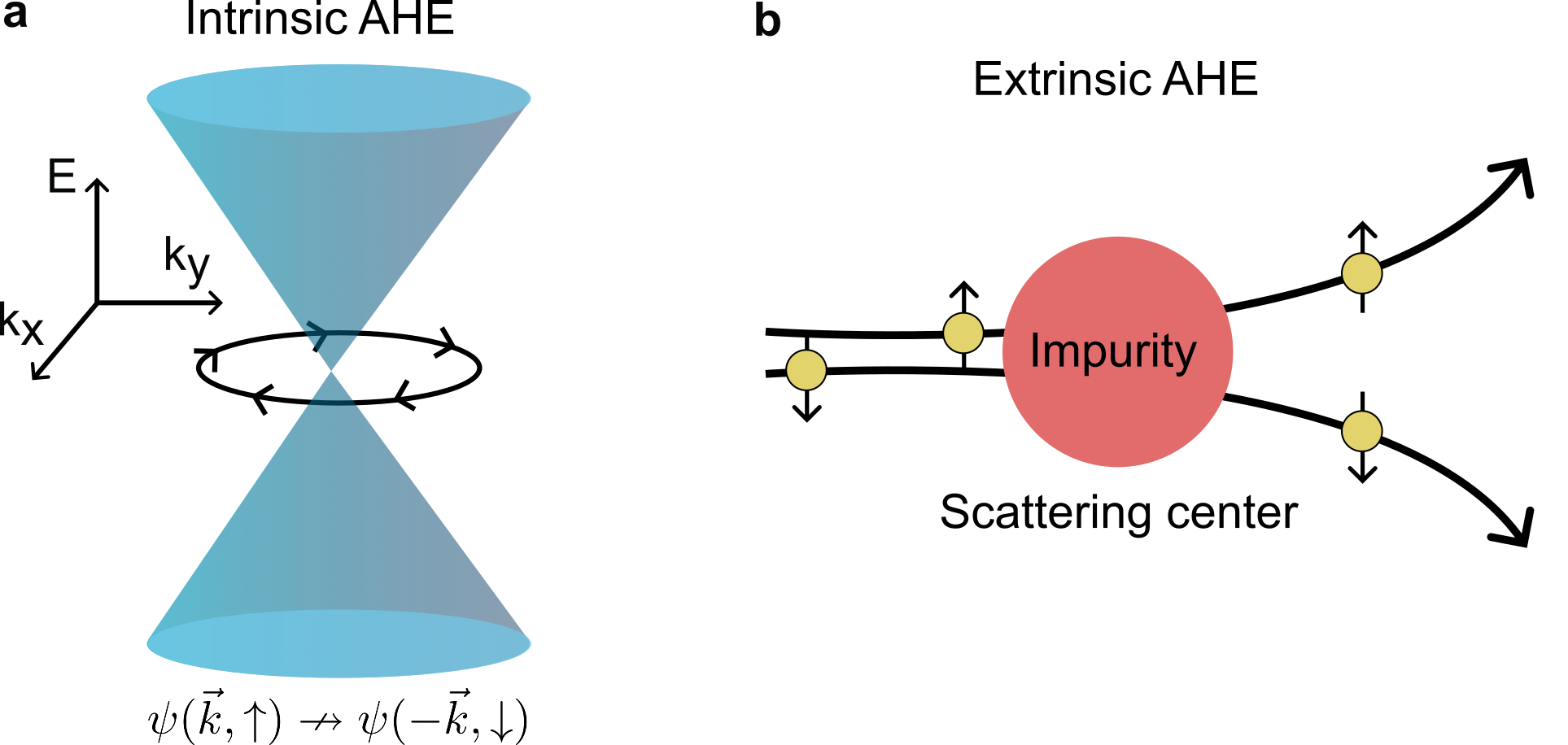}
		\endgroup
		\caption{(a) Intrinsic AHE arises from time reversal symmetry (TRS) breaking intrinsic to electronic structure. (b) Extrinsic AHE arises from impurity scattering mechanisms that break TRS.}\label{fig:1}
	\end{figure}
	
	\begin{figure*}[!thb]
		\centering
		\includegraphics[]{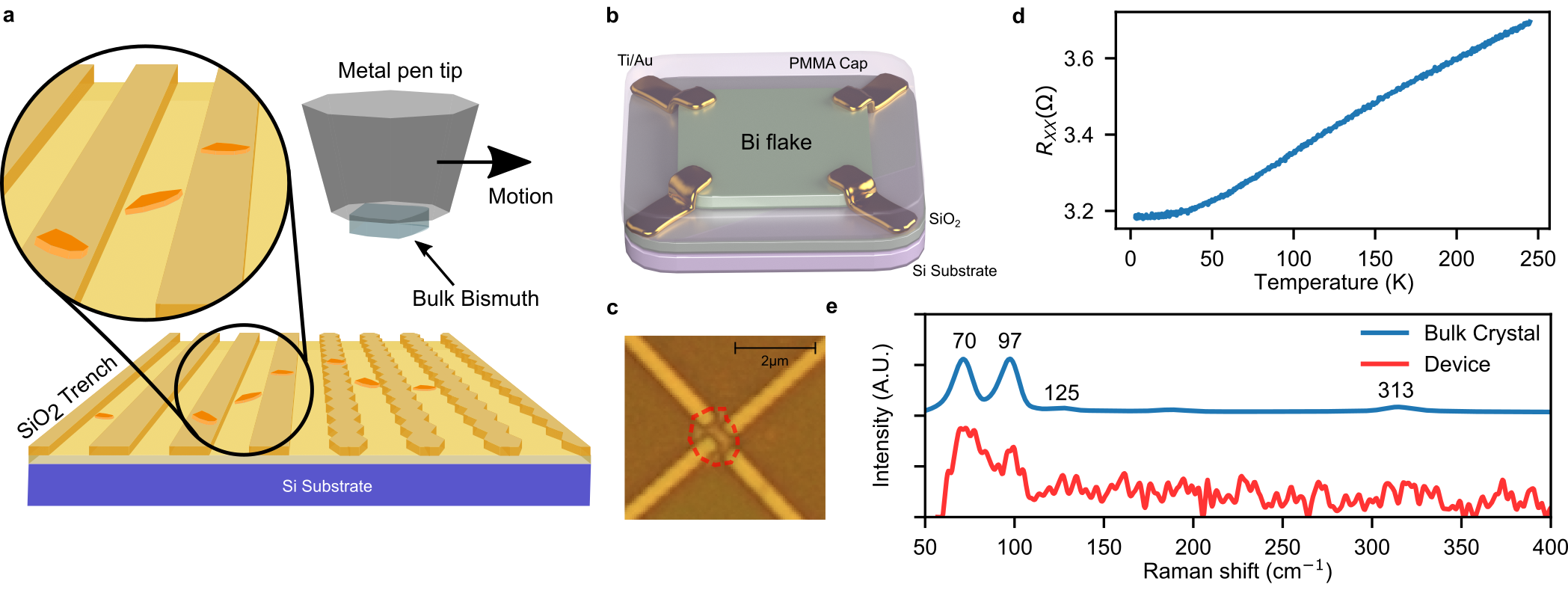}
		\caption{(a) Mechanical exfoliation of bismuth by grating bulk bismuth crystal against SiO$_2$ micro-trench structures. (b) Schematic of the device and (c) optical microscope image of the fabricated device in van der Pauw configuration where the red dashed lines indicate the perimeter of the flake as confirmed by AFM (see \textit{Supplemental Information}). (d) Four-point resistance as a function of temperature. (e) Raman spectroscopy of the device compared to its bulk counterpart. Note that 70 and 97 cm$^{-1}$ are Stokes shifts for pure bismuth whereas bismuth oxide $\beta$-Bi$_2$O$_3$ has a Raman peak at 313 cm$^{-1}$ which was not observed for the device. Raman spectra were normalized to the bismuth peak of 70 cm$^{-1}$.}\label{fig:2}
	\end{figure*}
	
	\par \textit{Introduction.---}The Anomalous Hall Effect (AHE) was discovered by Edwin Hall only a year after his discovery of the ordinary Hall Effect. However, unlike the classical Hall effect that was immediately rationalized, the mechanism for AHE remained the subject of debate for nearly a century. Today, it is believed that AHE has two types of contributions~\cite{AHEreview,AHEreview2}: an \textit{intrinsic}, scattering-free, mechanism originally proposed by Karplus and Luttinger~\cite{KarplusLuttinger} that can also be reconciled with the presence of a Berry curvature (see \cref{fig:1}a), as well as scattering-dependent mechanisms known as \textit{extrinsic} contributions (see \cref{fig:1}b). Despite these advances in the 1950s, AHE remains poorly understood in certain systems such as, for example, spin glasses. Notably with the recent emergence of topologically non-trivial band structures, the study and understanding of AHE in these materials led to a renaissance of interest in the topic. 
	
	\par Bismuth is the heaviest group V element ($Z=83$) and has therefore a very large atomic spin-orbit coupling (SOC)\cite{SOC,Hofmann}. However, due to inversion symmetry in the bulk Bi crystal, the bulk itinerant carriers do not experience SOC, in contrast to the surface state carriers. Bismuth in the bulk form has been extensively studied and has a long history of exhibiting celebrated transport properties. Remarkably, effects such as the Shubnikov-de Haas (SdH) effect, the de Haas-van Alphen effect and the Nernst–Ettingshausen effect were first discovered in bismuth. Recently, in part due to the discovery of graphene, interest in bismuth has been revived for its predicted properties in thin layered structures and in its 2D form known as bismuthene. For instance, a semimetal to semiconductor transition is predicted to occur when its thickness reaches 30 nm or less \cite{quantumEffectSsandomirskii,smsc,ultrathinBi}, but finding such transition has been proven to be experimentally difficult due to fabrication challenges and the potentially non-trivial contribution of surface states \cite{Hofmann}. Yet, quantum confinement is important in	understanding transport properties of films of thickness up to about 100 nm\cite{Xiao_2012,Marcano_2010,Jiang_2022}. Furthermore, advances were made in isolating the single layer allotrope, bismuthene, and it was successfully grown for the first time on a SiC substrate in 2017 \cite{bismutheneSHE}, albeit its transport properties still await to be unraveled experimentally. More recently, there has been a widespread interest in bismuth as there is substantial evidence for it to be a higher order topological insulator (HOTI)~\cite{HOTI1,HOTI2} and to host intrinsic superconductivity where the transition occurs below $T_c\approx0.5$ mK~\cite{SC}. As such, this makes bismuth an excellent candidate to study AHE in a system that is topologically non-trivial.
	
	\par If time-reversal symmetry (TRS) is broken in bismuth, then one would expect it to manifest the intrinsic AHE~\cite{AHEreview,AHEreview2}. For instance, the work of Y. Hirai \textit{et al.} on 30 nm bismuth film experimentally demonstrated that circularly polarized light can open a gap in the bismuth's band structure, leading to time-reversal symmetry breaking which then leads to an AHE~\cite{AHEbismuthlight}. Moreover, in a model proposed by Haldane in 1988, he argues that TRS breaking on a honeycomb lattice can lead to a Quantum Anomalous Hall Effect (QAHE)~\cite{haldane}. QAHE was first observed by Chang \textit{et al.}~\cite{QAHEfirst} in 2013, and in 2018 A. Young \textit{et al.}~\cite{QAHEtblg} experimentally found QAHE in twisted bilayer graphene where TRS was broken when the interlayer twist angle in the moiré pattern is $\theta\approx1.1^\circ$. Here, we report on our work in sub-100 nm bismuth films whereby an unambiguous signature of the AHE was observed in electronic transport measurements. Bismuth is known to be a diamagnetic material and as such the manifestation of AHE requires the breaking of TRS which is to our knowledge unexpected.  Our observation of the AHE in pure bismuth suggests that a hidden mechanism must be responsible for the TRS breaking. This discovery is not only important to further understand the already extensively studied properties of bismuth, but also to extend our comprehension of the AHE.
	
	\par \textit{Method.---} We developed a novel technique using micro-trench structure to mechanically exfoliate thin bismuth flakes~\cite{methodpaper}. The micro-trench structure was prepared by etching a SiO$_2$ thermal oxide layer above a degenerately doped silicon substrate, effectively turning it to a mechanical file. As shown in \cref{fig:2}a, bulk bismuth crystal, with its orientation carefully chosen to be the (1,1,1) surface, was attached to the tip of a metal pen. By grating the bismuth crystal against the micro-trench file, thin flakes of bismuth were obtained and found to be as thin as $\sim$10  nanometers~\cite{methodpaper}. Such mechanical exfoliation provides a way to obtain ultra-thin bismuth flakes in a clean and controlled environment. In particular, compared to the recently reported exfoliation methods using liquid sonication \cite{liquidex1,liquidex2}, water and oxygen that are detrimental to electronic properties can be avoided. While molecular beam epitaxy (MBE) techniques can produce films down to 3 nm \cite{dresselhaus}, our method is far more straightforward to prepare high quality flakes down to comparable thicknesses.
	
	\par Thin bismuth flakes with average thicknesses of 29 to 69 nm were obtained and characterized by atomic force microscopy (AFM), see \textit{Supplemental Information}. Note that the flakes have height variations, and only the average height is quoted here. Ti/Au contacts were deposited \textit{via} electron beam lithography (EBL) and electron beam vapor deposition. Note that the environment was carefully controlled with all fabrication steps performed in a vacuum, or in a nitrogen-filled glovebox. Lastly, a polymethyl methacrylate (PMMA) capping layer was spincoated for protecting the bismuth flake against oxidization. An optical image of a 68 nm device (see \textit{Supplemental Information}) fabricated in the van der Pauw (vdP) geometry is shown in \cref{fig:2}c, with the schematic shown in \cref{fig:2}b. Two other devices fabricated in a comb geometry are shown in the \textit{Supplemental Information}. Note that due to the small size of the flake ($\sim 1 \times 1$ \si{\micro\meter}), as we will discuss below, the ohmic contacts are subject to misalignments and hence mixing of electronic transport components is to be expected. 
	
	\par The chemical nature of bismuth flakes was confirmed \textit{via} Raman spectroscopy with a Bruker Senterra confocal Raman equipped with a 785 nm laser. The flake was compared to its bulk counterpart of the same crystal as a benchmark. The small Raman signal of the flake is due to its small size, however it was still possible to observe the 70 cm$^{-1}$ and 97 cm$^{-1}$ Raman shift peaks associated with pure bismuth~\cite{ramanBi1,ramanBi2}, see \cref{fig:2}e (red). Additionally, the most common type of bismuth oxide typically formed at lower temperatures ($\lesssim$300$^\circ$C), $\beta$-Bi$_2$O$_3$, was observed in the bulk crystal (blue) at 125 cm$^{-1}$ and 313 cm$^{-1}$~\cite{ramanBioxide1,ramanBioxide2,ramanBioxide3} but not in the exfoliated flake. 
	
	\par The temperature dependence of the resistance of the van der Pauw device is plotted in \cref{fig:2}d from 3 to 250 K, and is consistent with previously reported values for bulk semi-metallic bismuth~\cite{kukkonen,hoffman} as well as thin films down to 500 nm~\cite{partin}.
	
	\par \textit{Results and Discussions.---} The small size and the limitation imposed by EBL resulted in the van der Pauw contacts to be deposited relatively close to each other, as can be seen in \cref{fig:2}c. Since the contacts are not exactly at the corners as depicted in \cref{fig:2}b, the longitudinal (XX) and Hall (XY) resistances are expected to be mixed to some degree in every probe configuration. To overcome this mixing, we use the Onsager symmetrization to reconstruct the true longitudinal $R_{xx}$ and Hall $R_{xy}$ resistances. According to the Onsager's reciprocity theorem~\cite{onsager}, inverting the current and voltage contacts in a linear system allows us to measure the transpose of the resistance tensor given by
	$$R(B)=\left(\begin{matrix}
	R_{xx} & R_{xy}\\
	-R_{xy} & R_{xx}\\
	\end{matrix}\right), $$
	where $B$ is the applied magnetic field. Consequently, by measuring the resistance in one configuration as well as its Onsager reciprocal, the true longitudinal $R_{xx}$ and Hall $R_{xy}$ resistances can be obtained  by respectively symmetrizing and anti-symmetrizing the two configurations with 
	\begin{align*}
	R_{xx} &= \frac{R+R'}{2},\\
	\text{and}\qquad R_{xy} &= \frac{R-R'}{2},
	\end{align*}
	where $R$ and $R'$ form an Onsager pair. For example, a four-point measurement configuration labeled ABCD (corresponding to probes I+/V+/V-/I-) would have its Onsager reciprocal with contact configuration BADC. 
	
	\begin{figure}[!h]
		\includegraphics[]{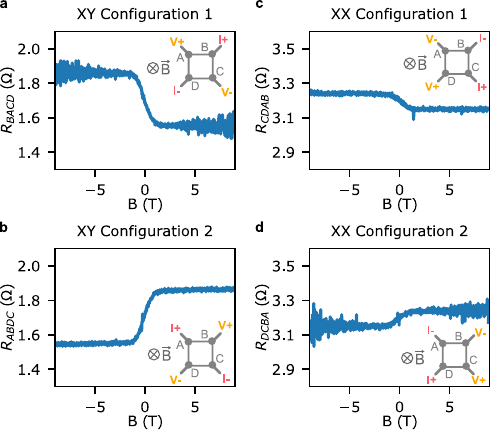}
		\caption{Four-point resistances of different probing configurations versus magnetic field $B$ (positive defined as pointed into the page) at 15 mK. (a) and (b) are in the XY configurations and form an Onsager pair, and (c) and (d) are in the XX configurations and are Onsager reciprocals. The contacts configurations are shown in the insets.}\label{fig:3}
	\end{figure}
	
	\begin{figure}[!h]
		\includegraphics[]{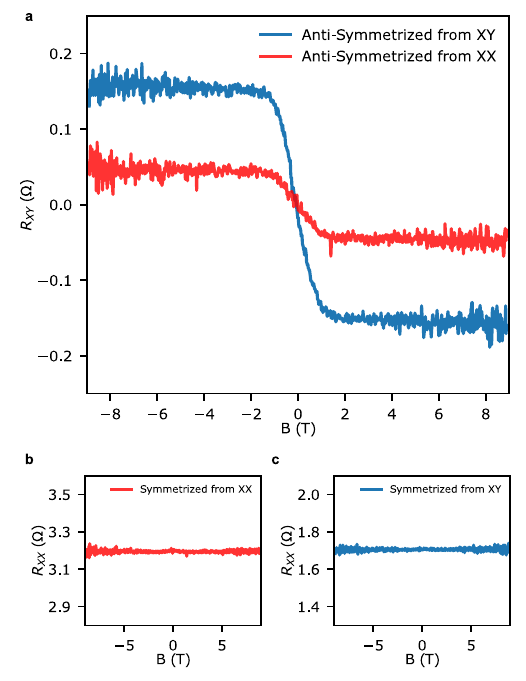}
		\caption{(a) $R_{xy}$ extracted from XX and XY Onsager pairs through anti-symmetrization, showing the Hall anomaly. (b) and (c) are the extracted $R_{xx}$ from the same Onsager pairs.}\label{fig:4}
	\end{figure}
	
	\par The resistances versus magnetic fields (with positive field defined as pointed into the page, see \cref{fig:3} insets) were measured at 15 mK and are shown in \cref{fig:3}. We stress that mixing of the $R_{xx}$ and $R_{xy}$ was inevitably observed, however \cref{fig:3}a and \cref{fig:3}b are XY configurations that maximize the Hall signal and that are also Onsager reciprocals to one another. Similarly, \cref{fig:3}c and \cref{fig:3}d are the XX configurations that optimize the XX signal while also capturing a mixed Hall signal. The corresponding ideal probe configurations in the van der Pauw geometry are shown in the insets of \cref{fig:3}. 
	
	\par Figure \ref{fig:4} summarizes our main result: the extracted true Hall and longitudinal resistances as a function of the magnetic field $B$. Figure \ref{fig:4}a shows the anti-symmetrized signal from the XY configurations, and as expected it is greater than the anti-symmetrized signal from the XX configurations. Similarly, the true longitudinal signal symmetrized in a similar fashion is shown in \cref{fig:4}b and \cref{fig:4}c for the XX and XY configuration pairs, respectively. In particular, both true $R_{xx}$ are constant as a function the magnetic field up to $\pm 9$ T. The resistance values are 3.2 \si{\ohm} (extracted from XX) and 1.7 \si{\ohm} (extracted from XY). The difference in the extracted resistances is attributed to the contacts that are not equidistant from one another, as the ratio of the $R_{xx}$ resistances is approximately equal to the ratio of the distances between the voltage probes.
	
	\par From the Hall response shown in \cref{fig:4}a, careful inspection of the data shows the presence of a very small slope in the saturated regime of the Hall signal.  In particular, for the XY configuration's saturated  high-field region ($|B|\ge 2$ T), a fit to the linear slope yields the value  $-0.0010(4)$ \si{\ohm/\tesla}, and similarly a linear fit to the XY configuration's low field region ($|B|\le0.5$ T) yields a slope of $-0.174(3)$ \si{\ohm/\tesla}, see \textit{Supplemental Information} for the detailed extraction of these values. While the negative values found for both linear Hall signals support the electronic transport being hole-dominated, it is unfortunately not possible to reliably extract carrier densities and mobilities because the constancy of the $R_{xx}$ shown in \cref{fig:4}(a) and (b) is uncharacteristic of the multi-carrier model. Bismuth is known to host  both  electron and hole pockets in the bulk~\cite{electronholebismuth,densitymobilitybismuth}, as well as in thin films~\cite{densitymobilitybismuthfilm,qconfinement,magnetotransportbismuth}. If both carriers were present, $R_{xx}$ can only saturate at high fields if the densities of electrons and holes were unequal, \textit{i.e.} $n\ne p$. The characteristic field for this saturation is given by \cite{magnetoRBi} $$B^\star = \dfrac{n\mu_p+p\mu_n}{|n-p|\mu_n\mu_p},$$ where $n$ and $p$ are electron and hole densities, and $\mu_n$ and $\mu_p$ are electron and hole mobilities. Note that the reduced Hall slope observed at high fields in \cref{fig:4}(a) implies that $|n-p|\ll 1$ which leads to a very high saturation field for $R_{xx}$. Under such saturation, $R_{xx}$ would have a magnetic field dependence $\propto B^2$, and if $R_{xy}$ is dominated by a low density, high mobility carrier at low fields, it would have a magnetic field dependence $\propto B^3$. Neither of these are observed in our work. Carrier density in bismuth can also have a field dependent effect \cite{zhu1, magnetoRBi, iwasa}, but the linearity of $R_{xy}$ with $B$ at high field implies that density variation is not responsible for the observations made here. Moreover, the elongated Fermi surface pockets known for bulk bismuth \cite{Hofmann} can strongly affect the transport behavior. That being said, the multi-carrier model discussed above is valid as long as the Fermi surfaces are closed. Meanwhile, the possibility of an open orbit in the Fermi surface can be excluded because it is inconsistent with the saturation of $R_{xx}$ at high field as observed in our experiments. All things considered, we cannot reconcile the multi-carrier model with the Hall and magnetotransport data presented in this work. Furthermore, in contrast to bulk bismuth where  quantum oscillations can be observed well under 1 T~\cite{sdhbismuth}, in our case  we did not observe any SdH oscillations in the longitudinal resistance. As we will discuss below,  this unexpected behaviour was nevertheless observed previously in a 100 nm thick film grown by MBE \cite{partin}.
	
	\par Using the device geometry and the longitudinal resistance measured in the absence of a magnetic field, the conductivity is calculated to be $\sigma_{xx}\sim 10^5$ $(\si{\ohm\centi\meter})^{-1}$ which places the bismuth device in a good-metal regime dominated by scattering-independent mechanisms~\cite{AHEreview}. It is indeed expected for bismuth to manifest the intrinsic AHE~\cite{AHEreview,AHEreview2} because of its high SOC. However, TRS breaking is a necessary condition for both intrinsic and extrinsic AHE, and as such, SOC alone is not sufficient to explain the AHE signal observed here. Usually, intrinsic TRS breaking is achieved by ferro- or antiferromagnetism, but bulk bismuth is known to be the most diamagnetic element with a magnetic susceptibility value of $-1.66\times10^{-4}$ ~\cite{bismuthsusceptibility}. Consequently, unless a magnetic transition would occur as the thickness of bismuth is reduced, TRS breaking must originate from elsewhere. Interestingly, in their growth and characterization of bismuth thin films by MBE, Partin \textit{et al.} found a Hall resistivity and magnetoresistivity nearly identical to our data for their thinnest 100 nm bismuth film (see Fig. 6 of~\cite{partin}).  Specifically,  their magnetoresistivity in the 100 nm film was found to be independent of the magnetic field and the Hall resistivity shows a similar AHE behaviour with a saturation near 1 T, as is observed in our work. Importantly, the work of Partin \textit{et al.} found more conventional behavior for films thicker than 100 nm, leading to the observation of SdH oscillations which is absent in the 100 nm film even for magnetic fields as high as 17 T.  The complete featureless and flat trend observed in their and our works for bismuth of similar thicknesses remains an intriguing mystery.
	
	\par More recently, B. C. Camargo \textit{et al.} claimed to have observed the AHE in bulk bismuth but were unable to find the source of the TRS breaking either~\cite{bismuthbulkAHE0}. A portion of their work was dedicated to eliminate magnetic contamination and superconductivity, and it was concluded that the AHE may arise from the topologically non-trivial surface or hinge states rather than in the bulk. Such arguments align with our observation of the AHE in thin devices with average flake thicknesses of 68 nm (main text), 29 nm and 69 nm (see \textit{Supplemental Information}). Strikingly, these two other devices were fabricated in a comb geometry, yet the four-probe resistances demonstrated a magnetic field response similar to that of AHE measured in the vdP device presented in the main section of our article. While the longitudinal resistances of the vdP and comb devices were different (2 to 32 \si{\ohm}), the anomalous Hall responses were unexpectedly similar (0.1 to 0.4 \si{\ohm}), hinting that the observed AHE does not arise from the bulk, but instead from the surface. The resistance's temperature dependence which is consistent with that of the bulk bismuth down to 500 nm also supports this observation as it suggests that the bulk is measured simultaneously with the surface or hinge states.
	
	\par Finally, in addition to a non-trivial topology in bismuth~\cite{HOTI1,HOTI2,bismutheneSHE,bismuthtopo}, there are also other prospect origins for spontaneous TRS breaking. These include band-flattening as in the case of twisted bilayer graphene, orbital magnetism of Dirac electrons that could arise in bismuth, and strain-induced band distortion~\cite{bismuthstrain,grapheneflatband,graphenestrain} that could all lead to a broken TRS. There is a long sequence of works supporting an intrinsic AHE in bismuth. From Conn and Donovan's work in the late 1940s which found traces of AHE~\cite{bismtuhbulkAHE1,bismuthbulkAHE2,bismuthbulkAHE0}, to the recent work by B. C. Camargo for the bulk~\cite{bismuthbulkAHE0}, as well as our own work in exfoliated bismuth that is consistent with Partin's MBE-grown 100 nm bismuth thin film, all these works point towards a {\it bona fide} intrinsic AHE occurring in bismuth with an unknown origin of broken TRS.  
	
	\par \textit{Conclusion.---} We successfully fabricated sub-100 nm bismuth devices and studied their electronic transport properties. Its measured Hall resistance is consistent with the Anomalous Hall effect despite bismuth being highly diamagnetic. Our results strongly suggest that there must exist a mechanism for breaking TRS in bismuth which needs to be further investigated and understood. In future works, we expect to explore the Hall anomaly at higher magnetic fields and at different field effect gate voltages to obtain a better understanding of its electronic band structure as well as a more profound comprehension of its predicted non-trivial topology. As such, and in spite of having been one of the most extensively studied materials since the 19$^\text{th}$ century, bismuth remains a fascinating elemental material that is yet to be fully understood.
	
	\section*{Acknowledgment}
	\par This work has been supported by the Canadian Institute for Advanced Research, the Canadian Foundation for Innovation, and the Fonds de Recherche du Québec Nature et Technologies. Sample fabrication was carried out at the McGill Nanotools Microfabrication facility and GCM Lab at Polytechnique Montréal. We would like to thank R. Talbot, R. Gagnon, and J. Smeros for technical assistance, M.-H. Bernier for helpful electron-beam lithography discussions, Ion Garate for useful discussions of the theory and finally Y. Wang for 3D drawings of the device. R. Allgayer acknowledges the financial support from the Vanier Canada Graduate Scholarship.

	\clearpage
	\newpage\pagebreak

\title{Supplemental Material: Anomalous Hall Effect in Thin Bismuth}
	\maketitle

	\onecolumngrid
\setcounter{figure}{0}
\renewcommand\thefigure{S\arabic{figure}}

\section{Transport Measurements}
\par The resistance of the thin bismuth device was measured with a voltage pre-amplifier using a quasi-DC technique at a frequency of 17.777 Hz and with an excitation current of 100 nA. The low-temperature measurements were performed in a Blufors BF-LD250 dilution refrigerator fitted with a 9 T magnet.

\section{Temperature Dependence}
\par Given that there are four different configurations used in the Onsager symmetrization, we show the resistance versus temperature traces for all of them for completeness. The $R_{xx}$ and $R_{xy}$ are shown in Fig. S1(a) and (b), respectively. Note that the trace for XX2 was shown in the main text of the manuscript.

\begin{figure}[!h]
	\centering
	\includegraphics[width=1\textwidth]{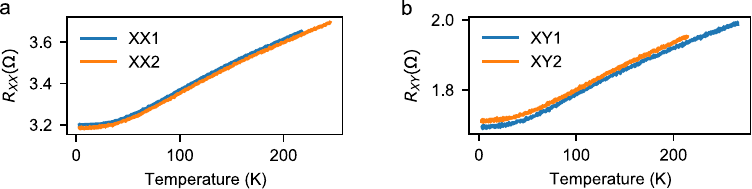}
	\caption{Temperature dependence of the resistance for (a) $R_{xx}$ Onsager pairs and (b) $R_{xy}$ Onsager pairs.}
\end{figure}

\section{Optical Images and AFM Scans}

\par The optical image before contact patterning is shown below in Fig. S2(a). The corresponding AFM scan is shown in Fig. S2(b). A profile is selected across the flake where the height versus position graph is shown in Fig. S2(c). Note that the latter shows a non-negligible height variation, and we found an average height of $68\pm47$ nm. Despite this height variation, note that we have observed similar AHE in our comb devices (see below) of different thicknesses, and as such we do not expect the unevenness to play an important role in the soundness of our conclusion. %$68\pm47$ nm

\begin{figure}[!h]
	\centering
	\includegraphics[width=1\textwidth]{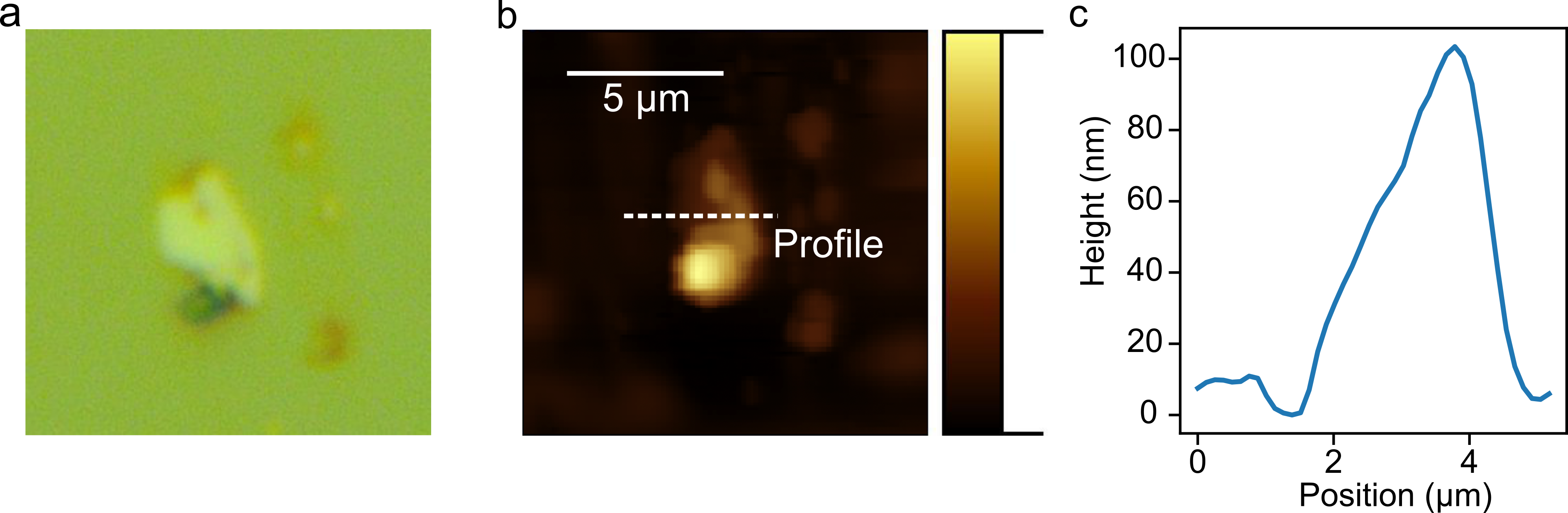}
	\caption{(a) Original optical image of the flake. (b) The corresponding AFM scan of the same flake. (c) The height profile as selected in the AFM scan.}
\end{figure}

\section{Additional Comb Devices}

\par We have observed the Anomalous Hall Effect in two other devices which we label device A and device B. These two devices were fabricated in a comb geometry rather than in a van der Pauw geometry as for the device presented in the main text. We observed the AHE in these two comb devices because of the mixing of transport components due to imperfections in the contacts. The optical images of the devices are shown below in Fig. S3. The thicknesses for devices A and B are $69\pm20$ nm and $29\pm19$ nm, respectively. The resistances are 19 \si{\ohm} and 32 \si{\ohm} for devices A and B, respectively. 

\begin{figure}[!h]
	\centering
	\includegraphics[width=0.6\textwidth]{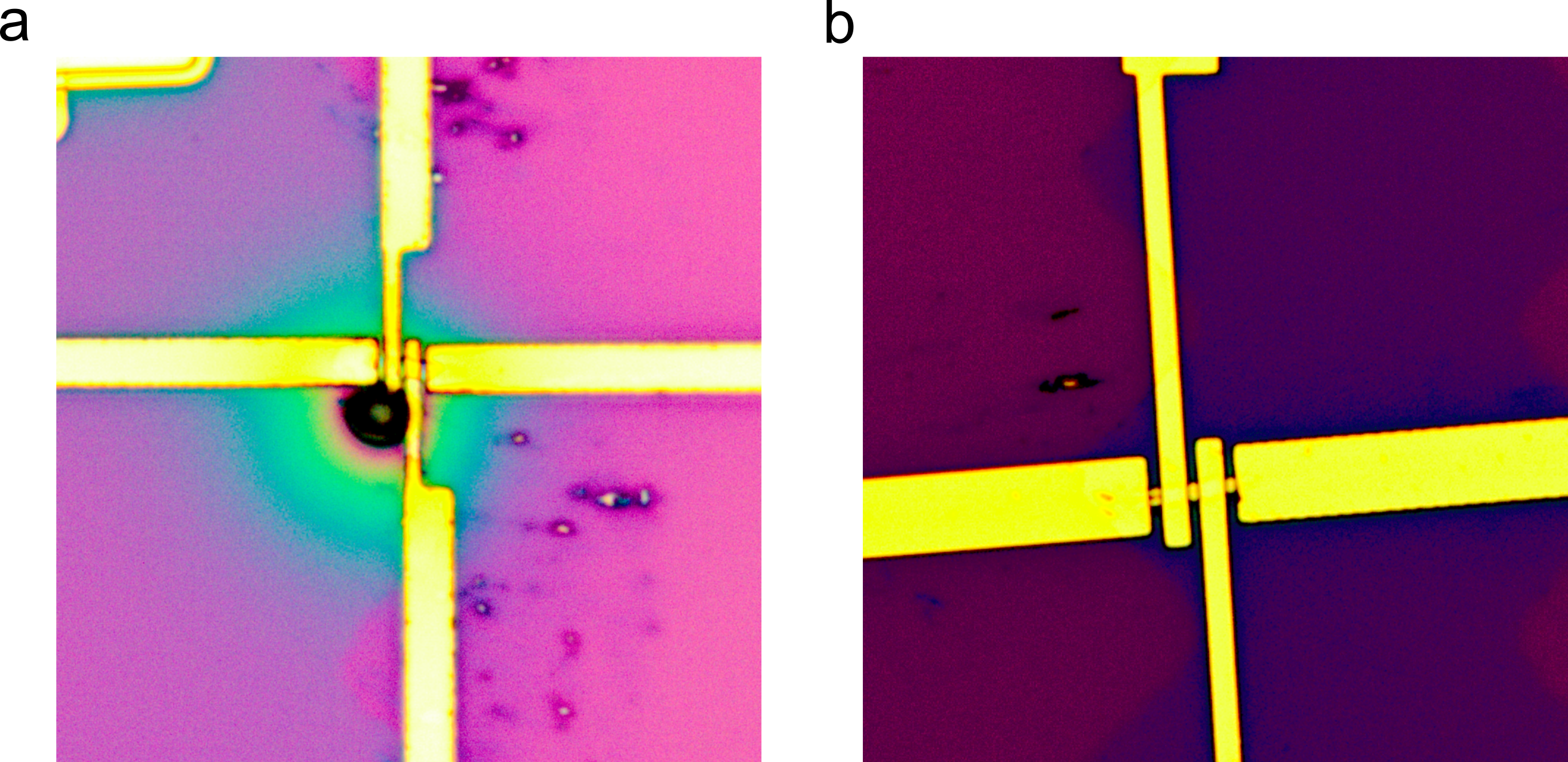}
	\caption{(a) 69 nm thin device A and (b) 29 nm thin device B both patterned the comb geometry. Note that the black dot in (a) is likely to be a particle of insulating SiO$_2$ left on the surface after cleaving the substrate for packaging.}
\end{figure}

\par For device A, we present the raw measured four-point resistance versus the magnetic field, shown below  in Fig. S4.

\begin{figure}[!h]
	\centering
	\includegraphics[width=0.6\textwidth]{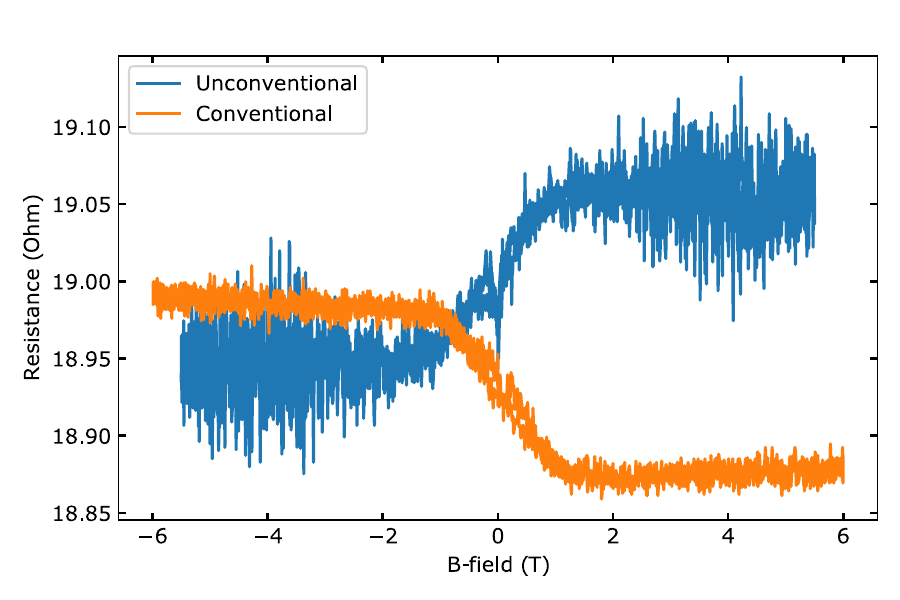}
	\caption{Four-point resistance versus the magnetic field for device A.}
\end{figure}

\par The conventional configuration is the usual I+/V+/V-/I- configuration where the current contacts are on the outside of the comb geometry and the voltage probes are in the center. In contrast, the unconventional configuration is obtained by inverting I+ with V+ and I- with V-. This effectively becomes the Onsager reciprocal of the conventional setup. Although the AHE is clearly observed,  the two curves do not cross at the origin which we attribute tentatively to the non-ideal configuration for Hall response used here.

\begin{figure}[!h]
	\centering
	\includegraphics[width=0.6\textwidth]{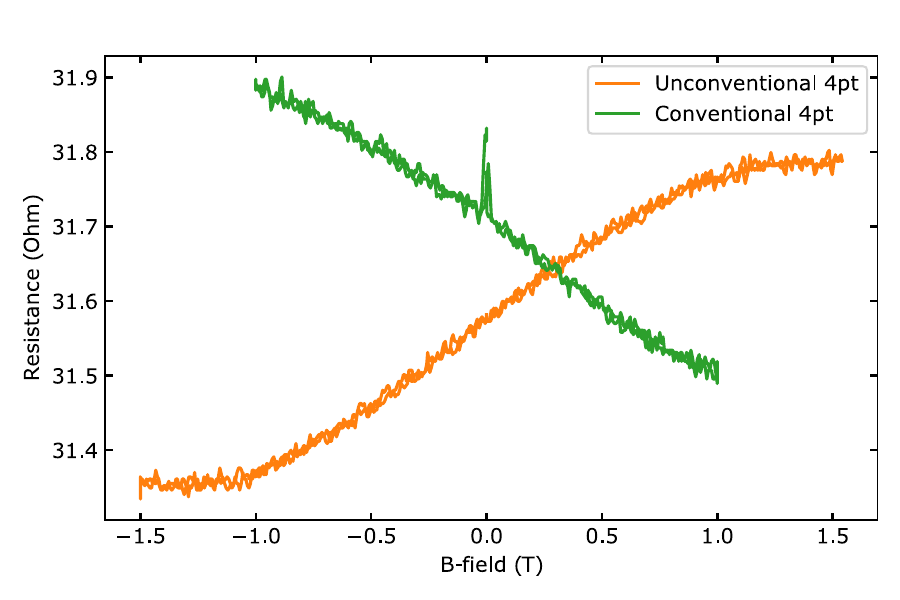}
	\caption{Four-point resistance versus the magnetic field for device B.}
\end{figure}

\par  In Fig. S5 below we show the resistance dependence on the magnetic field for device B. Similarly to device A, the AHE is clearly observed despite a vertical offset most likely due to the non-ideal configuration for measuring a Hall response.

\section{Linear Fits of the Anomalous Hall Response}
\par We perform linear fits on the anti-symmetrized data shown in the main text of the manuscript.  The high-field region is chosen to be $|B|\ge 2$ T and the low field region is chosen as $|B|\le0.5$ T. The corresponding slopes are shown in Fig. S6.

\begin{figure}[!h]
	\centering
	\includegraphics[width=0.6\textwidth]{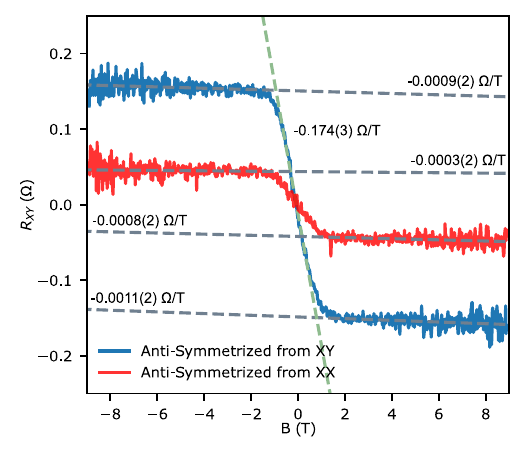}
	\caption{Linear fit for the high and low field regions of the AHE data shown in the main text of the manuscript.}
\end{figure}

\end{document}